\documentclass[%
 nofootinbib,
 amsmath,amssymb,
 prl,
 twocolumn,
 notitlepage
]{revtex4-1}

\usepackage{amsmath,amssymb,amsfonts}
\usepackage{amsthm}
\usepackage{algorithm}
\usepackage{algpseudocode}
\usepackage{algorithmicx}
\usepackage{graphicx}
\usepackage[font=footnotesize,labelfont=bf,justification=raggedright]{caption}
\usepackage{textcomp}
\usepackage{braket}
\usepackage{bm}
\usepackage{color}
\usepackage{hyperref}

\begin{document}
\title{Entanglement Properties of Quantum Superpositions \\of Smooth, Differentiable Functions}

\author{Adam Holmes}
\email{adam.holmes@intel.com}
\email{adholmes@uchicago.edu}
\affiliation{Intel Labs, Hillsboro, OR 97124, USA}
\affiliation{The University of Chicago, Chicago, IL 60615, USA}

\author{A. Y. Matsuura}
\affiliation{Intel Labs, Hillsboro, OR 97124, USA}

\date{\today}

\begin{abstract}
We present an entanglement analysis of quantum superpositions corresponding to smooth, differentiable, real-valued (SDR) univariate functions. SDR functions are shown to be scalably approximated by low-rank matrix product states, for large system discretizations. We show that the maximum von-Neumann bipartite entropy of these functions grows logarithmically with the system size. This implies that efficient low-rank approximations to these functions exist in a matrix product state (MPS) for large systems. As a corollary, we show an upper bound on trace-distance approximation accuracy for a rank-2 MPS as $\Omega(\log N/N)$, implying that these low-rank approximations can scale accurately for large quantum systems.
\end{abstract}

\maketitle

%\section{Introduction}
\label{sec:introduction}
The promise of quantum computation is the ability to solve problems exponentially faster than we can today with classical computers. However, the performance of many quantum algorithms depends on the ability to load classical data efficiently and accurately into quantum states. For example, this capability is necessary for quantum computers to be viable for performing machine learning with large classical training data sets \cite{harrow2009quantum,childs2017quantum,lloyd2014quantum,rebentrost2014quantum,wiebe2012quantum} and Monte Carlo calculations that compute expectation values of functions over classical probability distributions \cite{harrow2009quantum,woerner2019quantum,stamatopoulos2019option,montanaro2015quantum}. In both of these cases, preparing the state corresponding to the data is vital to preserving the quantum speedup present in the remainder of the algorithm. While in general, state preparation is an exponentially hard problem \cite{plesch2011quantum}, it has previously been shown empirically that certain families of quantum states can be prepared efficiently and with high precision using linear-depth circuits, and algorithms have been developed that require only linear compute time to generate these linear-depth circuits for real-valued smooth, differentiable (SDR) functions \cite{holmes2020efficient}. The practical utility of this approach depends upon whether it is scalable to large qubit systems. 

In this paper, we prove analytical upper bounds on the entanglement requirements for quantum state superpositions of discretized SDR functions, and thereby assess and demonstrate scalability of state preparation encoding methods using matrix product state (MPS) data structures. %The present work also builds upon earlier work \cite{schuch2008entropy} in which bounds were provided establishing matrix product states simulation power for low entanglement quantum states. 
These have been shown to be powerful for simulating correlated one-dimensional systems \cite{vidal2003efficient,vidal2004efficient}, and bounds have been established showing domains over which MPS encodings remain efficient data structures \cite{schuch2008entropy}. In the present work, we prove that superpositions corresponding to discretized SDR functions require entanglement that grows logarithmically in system size, thereby also showing that they can be accurately simulated with matrix product states as is  shown in \cite{schuch2008entropy}. 

%\section{Results}
\label{sec:proof}
\newtheorem{theorem}{Theorem}
\newtheorem{lemma}{Lemma}
\newtheorem{corollary}{Corollary}
%We study the von-Neumann entropy of quantum superposition states that correspond to discrete evaluations of smooth, differentiable, real-valued (SDR) univariate functions. 

The proof structure is quite simple. We construct a polynomial approximation to the target quantum state% up to specified accuracy, 
and bound the entropy of an encoding of this polynomial into a corresponding MPS. The entropy of this encoding is upper bounded by the bond-dimension of the MPS, which is used to complete the proof. The main result is stated and proved in Theorem \ref{thm:main}, and relies on several lemmas which are stated first.

The first lemma bounds the pointwise approximation accuracy of polynomial function approximations in terms of the domain of the function $D$, the derivative of the function $\gamma$, and the polynomial approximation degree $k$. 
\begin{lemma}
\label{lemma:pointwise}
Let $D \subseteq \mathbb{R}$ be a closed bounded interval and assume that $f$ is a function with bounded derivatives that satisfies, for some $C_f, \gamma_f \geq 0$
\begin{align}
    \Vert f^{(n)} \Vert_{\infty} \leq C_f \gamma_f^n n! \quad\  \forall n \in \mathbb{N}_0
\end{align}
then a polynomial $g$ with support over the same domain $D$ requires a degree $p$ to achieve $\ell_{\infty}$ pointwise accuracy $\Vert f-g\Vert_{\infty}\leq\varepsilon$ as:
\begin{align}
    p = \log_{\alpha}\big(1/\varepsilon \big) + C
\end{align}
for constant $C$ and $\alpha = 1 + \frac{2}{\gamma_f |D|}$.
\begin{proof}
Lemma 3.13 of  \cite{borm2005approximation} shows that for the appropriate domains $D$ and functions $f$, we have that, for all $k \in \mathbb{N}_0$
\begin{align}
    &\min_{g \in \mathcal{P}_k} \Vert f - g \Vert_{\infty,D} \leq \frac{C_f 4 e (1+\gamma_f |D|)(k+1)}{\big(1+\frac{2}{\gamma_f |D|}\big)^{(k+1)}}
\end{align}
where the family $\mathcal{P}_k$ is the family of polynomials of degree $k$. This is used in \cite{grasedyck2010polynomial} to prove the statement above.
\end{proof}
\end{lemma}
The next lemma is a simple upper bound on the $\ell_1$ norm of quantum states, or normalized vectors $\boldsymbol{v} \in \mathbb{C}$. 

\begin{lemma}
\label{lemma:max_onenorm}
Let $f$ be a discretized SDR function over a closed bounded interval $D \subseteq \mathbb{R}$ normalized such that $\Vert f \Vert_2 = 1$. Then we have:
\begin{align}
    \max_f \Vert f \Vert_1 \leq 2^{N/2}
\end{align}
This bound is tight, saturated by the uniform distribution as:
$f_i = 2^{-N}$ for all entries.
\begin{proof}
Apply the Cauchy-Schwarz inequality and use the all-ones vector $\textbf{1} = \{1\}_{i=1}^{2^N}$:
\begin{align}
    \Vert f \Vert_1 &= \Vert f \cdot \textbf{1} \Vert_1 \leq \Vert f \Vert_2 \Vert \textbf{1} \Vert_2\\
    &=1 \bigg(\sum_{i}^{2^N} 1\bigg)^{1/2}=2^{N/2}
\end{align}
We saturate this with the uniform distribution $u$ with $u_i = 2^{-N/2}$ for all $i$ as $\Vert u \Vert_1 = 2^N (2^{-N/2}) = 2^{N/2}$.
\end{proof}
\end{lemma}
The last lemma is a lower bound on the overlap, or inner product, of a quantum superposition of a discretized SDR function and an approximation to this state constructed with a fixed degree polynomial. This allows us to convert between the entrywise approximation error infinity-norm $\ell_{\infty}$ to a bound on the inner product or the two-norm $\ell_2$.

\begin{lemma}
\label{lemma:connect}
Let $f$ be a unit normalized discretization of a univariate SDR function $f(x)$ over a closed, bounded interval $D \subseteq \mathbb{R}$ covered by $2^N$ equidistant evaluation points, such that $\Vert f \Vert_{2} = 1$. Let $g$ be a unit vector pointwise approximation of $f$ on a closed, bounded interval $D$ such that $\Vert f-g \Vert_{\infty} = \varepsilon$.
Then
\begin{align}
    \langle f, g \rangle \geq 1 - \varepsilon^2 2^{N-1}
    \label{eq:lemma3}
\end{align}
\begin{proof}
Begin by considering an $\varepsilon$-pointwise approximation $g$ to $f$ without requiring $\Vert g \Vert_2 = 1$. In this case, we can set:
\begin{align}
    g_i \equiv f_i - \varepsilon \quad\ \forall i \in [1,2^N]
\end{align}
This choice for $g$ minimizes $\langle f, g \rangle$ to:
\begin{align}
    \langle f, g \rangle &= \sum_{i}^{2^N} f_i(f_i - \varepsilon)
    = \sum_i^{2^N} (f_i^2 - \varepsilon f_i)\\
    &= 1 - \varepsilon \sum_i^{2^N} f_i = 1 - \varepsilon \Vert f \Vert_1 \leq 1 - \varepsilon2^{N/2} 
    \label{eq:looser_bound}
\end{align}
where Lemma \ref{lemma:max_onenorm} has been used in equation \ref{eq:looser_bound}.
Consider any other $g'$ defined by changing some of the entries as:
\begin{align}
    g'_j = f_j + \varepsilon_j
\end{align}
for any $-\varepsilon < \varepsilon_j < \varepsilon$, with the only constraint that the entries follow $|f_j - g_j| \leq \varepsilon$ as dictated by the pointwise approximation error. We then have:
\begin{align}
    \langle f, g' \rangle &= \sum_{i}^{2^N} f_i(f_i + \varepsilon_i)
    = \sum_i^{2^N} (f_i^2 + \varepsilon_i f_i)\\
    &= 1 + \sum_i^{2^N}\varepsilon_i f_i \label{eq:min_g'}
\end{align}
Minimizing equation \ref{eq:min_g'} requires minimizing a summation over the distribution of $\varepsilon_i$, which is clearly minimized for $\varepsilon_i = -\varepsilon$ for all $i$. 

Requiring that $\Vert g \Vert_2 = 1$ will tighten the bound:
\begin{align}
    \Vert g \Vert_2 &= 1 = \sum_i g_i^2 = \sum_i (f_i + \varepsilon_i)^2\\
    &= \sum_i (f_i^2 + 2\varepsilon_i f_i + \varepsilon_i^2)\\
    &= \sum_i f_i^2 + 2\sum_j \varepsilon_j f_j + \sum_k \varepsilon_k^2
\end{align}
which because $|\varepsilon_i| \leq \varepsilon$ implies %by Lemma \ref{lemma:max_onenorm} 
that:
\begin{align}
    -2\sum_i \varepsilon_i f_i &= \sum_j \varepsilon_j^2 \leq 2^N\varepsilon^2\\
    \sum_i\varepsilon_if_i &\geq -2^{N-1} \varepsilon^2.
\end{align}
We can then tighten the bound in equation (\ref{eq:looser_bound}) to:
\begin{align}
    \langle f, g \rangle &= \sum_i f_i g_i = \sum_i f_i (f_i + \varepsilon_i)\\
    &= \sum_i f_i^2 + \varepsilon_i f_i = 1 + \sum_i\varepsilon_if_i\\
    &\geq 1 - 2^{N-1}\varepsilon^2
\end{align}
%for SDR function $f$ and approximation $g$ that satisfy:
%\begin{align}
%    \varepsilon \leq \frac{\Vert f \Vert_1 - 2f_{\text{min}}}{2^{N-1}}
%\end{align}
%where $f_{\text{min}} = \min_i f_i$, and we allow $g_i = f_i - \varepsilon$ for all $i \neq k$, and we set $g_k = f_i + \delta$ for the appropriate normalizing correction.
\end{proof}
\end{lemma}

\begin{corollary}
For a fixed \emph{error} $\delta$, the required pointwise error $\varepsilon = \Vert f - g \Vert_{\infty}$ is given by:
\begin{align}
    \varepsilon \leq \sqrt{\frac{\delta}{2^{N-1}}}
    \label{eq:corollary1}
\end{align}
\begin{proof}
This is a simple rewriting of \ref{eq:lemma3} with some fixed \emph{error} $\delta$ so that:
\begin{align}
    \langle f,g \rangle &\geq 1-\varepsilon^2 2^{N-1} \geq (1-\delta)\\
    \varepsilon^2&\leq \frac{1-(1-\delta)}{2^{N-1}}
\end{align}
\end{proof}
\end{corollary}

We now state the main result. By convention, we use the definition of the $K$-qubit reduced density matrix bipartite von-Neumann entropy as:
\begin{align}
    S(\rho_K) &= -\text{Tr}[\rho_K\log \rho_K]\\
    &= \sum_i |\lambda_i|^2 \log |\lambda_i|^2 \label{eq:vne_pure}
\end{align}
where $\rho_K$ is a $K$-qubit reduced density matrix, and equation \ref{eq:vne_pure} holds for pure state Schmidt decompositions of $\rho_K$ with $\lambda_i$ Schmidt coefficients. 

\begin{theorem}
\label{thm:main}
Let $\rho$ be the quantum superposition corresponding to the discretization of a univariate SDR function $f(x)$ over domain $D \subset \mathbb{R}$ covered by $2^N$ equidistant evaluation points. Then:
\begin{align}
    S_{\text{max}}(\rho_K) \leq \mathcal{O}(\log N)
    \label{eq:main}
\end{align}
where $\rho_K$ is any $K$-qubit reduced density matrix of $\rho$. 
\end{theorem}
This states that the maximum von-Neumann bipartite entropy of  a discretized SDR function as a quantum state grows as $\mathcal{O}(\log N)$. Unless explicitly specified otherwise, all logarithms are assumed to be base-2.
\begin{proof}
Begin by writing the $\varepsilon-$accurate pointwise approximation to $f(x)$ as $g$, such that $\Vert f-g \Vert_{\infty} = \varepsilon$. By Lemma \ref{lemma:pointwise}, we state the degree $p$ of a polynomial $g$ required to achieve pointwise accuracy $\varepsilon$ as:
\begin{align}
    p = \log_{\alpha}\big(1/\varepsilon \big) + C
    \label{eq:required_degree}
\end{align}
for constant $C$ and $\alpha = 1 + \frac{2}{\gamma_f |D|}$.  %as the minimum rank required of a tensor-train approximation to $f$. 
By Lemma \ref{lemma:connect} and the corresponding Corollary \ref{eq:corollary1}, we write the required error $\varepsilon^*$ for fixed error $\delta$ as:
\begin{align}
%    \varepsilon^* &\leq \frac{\delta}{\Vert f \Vert_1} \leq \delta 2^{-N/2}
    \varepsilon^* &\leq \sqrt{\frac{\delta}{2^{N-1}}} = \sqrt{\delta} 2^{-(N-1)/2}
    \label{eq:required_eps}
\end{align}
%where the second inequality holds by Lemma \ref{lemma:max_onenorm}. 
Plugging equation \ref{eq:required_eps} into equation \ref{eq:required_degree} for constant error $\delta \ll 1$ we have:
\begin{align}
    p &= \frac{\log \big(2^{(N-1)/2}/\sqrt{\delta}\big)}{\log \big(1+\frac{2}{\gamma_f|D|}\big)}\\
    &= \frac{\log \big(2^{(N-1)/2}\big) - \log \big(\sqrt{\delta}\big)}{\log (1+\frac{2}{\gamma_f|D|})}\\
    &= \frac{(N-1)-\log \delta}{2\log (1+\frac{2}{\gamma_f|D|})}
    \label{eq:approx_degree}
\end{align}

We now use the fact that a MPS encoding of a polynomial of degree $p$ requires a maximum bond-dimension $\chi \leq p+1$ \cite{grasedyck2010polynomial}. Because of this, and because the maximum achievable von-Neumann bipartite entropy of a rank-$\chi$ MPS is $\log_2(\chi)$ \cite{eisert2013entanglement}, we have, for constants $C_1, C_2$:
\begin{align}
%    S_{\text{max}}(\rho_K) &\leq \log(p+1)\\
%    &= \log \bigg(\frac{N/2}{\log (1+2/\gamma_f|D|)}+1\bigg)\\
%    &= \log \bigg(\frac{N/2 + \log (1 + 2/\gamma_f |D|)}{\log (1+2/\gamma_f|D|)}\bigg)\\
%    &= \log (N/2 + C_1)-\log (C_1)\\
%    &= \mathcal{O}(\log (N))
     S_{\text{m}}(\rho_K) &\leq \log(p+1)\\
    &= \log \bigg(\frac{(N-1)-\log \delta}{2\log (1+\frac{2}{\gamma_f|D|})}+1\bigg)\\
    &= \log \bigg(\frac{N-1-\log \delta + 2\log (1+\frac{2}{\gamma_f|D|})}{2\log (1+\frac{2}{\gamma_f|D|})}\bigg)\\
    &= \log (N - \log \delta + C_1)-C_2\\
    &= \mathcal{O}(\log N)   
\end{align}
as stated.
\end{proof}

Theorem \ref{thm:main} indicates that SDR function discretizations as quantum superpositions have entanglement entropy that scales logarithmically with the size of the system. As others have shown \cite{schuch2008entropy}, this implies efficient storage in a MPS data structure for all such functions. 

Additionally, there is only a weak %($\log \log$)
dependence on the maximum analytical derivative of the function: $\gamma_f$. As long as $\gamma_f$ grows sub-exponentially, then the function will remain efficiently stored by a polynomial number of parameters in a MPS.  %This implies that so long as %$\log \log \gamma_f = o(\log N)$, or equivalently $\log \gamma_f \leq \mathcal{O}(N)$, 

We can go farther and explore the accuracy of bounded bond-dimension ($\chi$) MPS approximations to discretized SDR function superpositions. To do so, we invoke the Fannes-Audenaert entropy bound \cite{audenaert2007sharp}, and use it to upper-bound the overlap between an SDR function discretization superposition and a fixed-$\chi$ MPS approximation to these states. This yields an upper limit to the ability of compressed MPS forms to accurately approximate these states.

\begin{corollary}
\label{thm:cor}
Let $f$ be the unit normalized exact discretization of a univariate SDR function $f(x)$ over domain $D \subset \mathbb{R}$ covered by $2^N$ equidistant evaluation points, such that $\Vert f \Vert_2 = 1$. Also, let the corresponding quantum state $\rho_f$ saturate the entropy bound in Theorem \ref{thm:main}: $S_{max}(\rho_{f,K}) \leq \mathcal{O}(\log N) = C_0 \log N$ for some constant $C_0$. Let $g$ be the unit normalized function corresponding to a rank-2 matrix product state approximation of $f$, denoted by $\sigma_2$. Then: 
\begin{align}
\Vert \rho - \sigma_2 \Vert_{tr} = \Omega \bigg(\frac{\log N}{N}\bigg)    
\end{align}
and for constants $A, B$:
\begin{align}
    \langle f, g\rangle &= \mathcal{O}\bigg(\sqrt{1-A^2\big( \log (N + B)/N\big)^2}\bigg)% \sqrt{1-\bigg(\frac{2\log N + C_2}{N C_1 }-1\bigg)^2}
    %&= \mathcal{O}\bigg(\frac{\log N}{N}\bigg)
\end{align}
\begin{proof}
We show this using the Fannes-Audenaert entropic bound \cite{audenaert2007sharp}.
\begin{align}
|S(\rho) - S(\sigma_2)| &\leq T \log (2^N-1)+H((T,1-T))\\
&\leq \frac{1}{2}\Vert \rho-\sigma_2\Vert_{\text{tr}}\log(2^N-1)+1
\end{align}
where $T = \frac{1}{2}\Vert \rho-\sigma_2\Vert_{\text{tr}}$ and $H$ is the binary Shannon entropy, upper bounded by 1.

By rearranging the bound, we see:
\begin{align}
   \Vert \rho-\sigma_2\Vert_{\text{tr}} &\geq \frac{2|S(\rho) - S(\sigma_2)|}{\log(2^N-1)}-1\\ 
%   &= \frac{2}{N}\bigg|\frac{\log (N/2 + C_1)}{\log (C_2)}-1\bigg|-1\\
   &= \frac{2}{N}\bigg|\frac{C_0\log (N - \log \delta + C_1)}{\log (C_2)}-1\bigg|-1\\
   &= \frac{2C_0\log (N - \log \delta + C_1-2\log C_2)}{N\log (C_2)}\\
   &= \Omega\bigg(\frac{\log N}{N}\bigg)
\end{align}
This holds in the limit of large systems $N$, $\log(2^N-1) \rightarrow N$, and with the maximum entropy saturating the upper bound provided by Theorem \ref{thm:main} and the maximum entropy of a rank-2 MPS as 1. $S(\rho) - S(\sigma_2)$ is assumed to be positive.

By exploiting the fact that these states are all pure states, we can convert the trace distance to fidelity \cite{nielsen2002quantum} and write:
\begin{align}
    \langle f, g \rangle &= \sqrt{1-\Vert f - g \Vert^2_{\text{tr}}}\\
%    &\leq \sqrt{1-\bigg(\frac{2(\log N/2 + C_1)}{C_2 N}-2\bigg)^2}
%    &\leq \sqrt{1-\bigg(\frac{2}{N}\bigg|\frac{\log (N - \log \delta + C_1)}{\log (C_2)}-1\bigg|-1\bigg)^2}\\
    &\leq \sqrt{1-\bigg(\frac{2\log (N - \log \delta + C_1 - 2\log C_2)}{N \log C_2 }\bigg)^2}\\
    &=\mathcal{O}\bigg(\sqrt{1-A^2\big(\log (N + B)/N\big)^2}\bigg)
    %&= \mathcal{O}\bigg(\frac{\log N}{N}\bigg) 
\end{align}
\end{proof}
\end{corollary}
Note that this is an upper bound on the trace distance between these two quantum states: the original, and a polynomial approximator. In the best cases, $S(\rho_{f,K})$ is bounded below $\log N$, in which case the approximator with constant entropy can competitively construct a high-accuracy state. These bounds imply that approximations can become exact for large system sizes.%In the worst case, the original state saturates the entropy of the bound in Theorem \ref{thm:main}, which is the more interesting case. Here we find that as the system size grows, the closer the upper bound gets to unit or exact solution. More precisely, for functions $f(x)$ with states $\rho_f$ that possess logarithmic maximum von-neumann entropy: $S_{max}(\rho_{f,K}) = \mathcal{O}(log N)$, we then have that: $\langle f, g \rangle \sim \frac{log N}{N}$, which approaches 

%\section{Results and Applications}
%\label{sec:results}
%\input{results.tex}

%\section{Discussion}
\label{sec:conclusion}
In conclusion, constructing quantum states that correspond to classical data is vital to preserving quantum speedup for applications that rely on large sets of classical input data. This work demonstrates that, as long as classical data sets are approximable by smooth, differentiable, real-valued functions, then the maximum bipartite von-Neumann entropy of the corresponding pure quantum state grows only logarithmically in the size of the system, or dataset. Equivalently, the required entanglement necessary to fully describe the dataset scales only logarithmically with the amount of data being considered. As data becomes more unstructured, constant factors in equation (\ref{eq:approx_degree}) begin to increase the required polynomial approximation degree, which in turn requires more entanglement to fully describe the data inside a quantum register.

\vspace{-1pt} 

Corollary \ref{thm:cor} shows that this logarithmic scaling actually indicates the existence of highly accurate, low-rank matrix product states that approximate these states, for large data sets. Because the Fannes-Audenaert bound is tight, this upper bound on the state fidelity of a rank-2 approximation is good evidence that approaches like that of \cite{holmes2020efficient} will scale for larger data sets. In fact, Corollary \ref{thm:cor} indicates that as long as the data is well-structured, the larger the data set becomes, the better the best rank-2 matrix product state approximation becomes. Based on this reasoning, the technique developed in \cite{holmes2020efficient} that seeks out this approximation is expected to scale efficiently and remain accurate for larger classical data sets and functions. 

\bibliographystyle{plain}
\bibliography{references}

\end{document}